# How Droplets Move on Surfaces with Directional Chemical Heterogeneities

David Feldmann, Bat-El Pinchasik*

D. Feldmann, B.- E. Pinchasik

Tel-Aviv University

School of Mechanical Engineering

Faculty of Engineering

Tel-Aviv, Israel

E-mail: pinchasik@tauex.tau.ac.il

**Keywords**: wetting, droplet roll-off, adhesion forces, biphilic surfaces, directional transport

**Abstract**

The nature of adhesion of droplets to surfaces is a long pending scientific question. With the evolution of complex surfaces, quantification and prediction of these forces become intricate. Nevertheless, understanding these forces is highly relevant for explaining liquid transport in nature and establishing design guidelines for manmade interfaces. Here, it is shown that the adhesion of droplets is highly sensitive to the directionality of chemical heterogeneities. Asymmetric biphilic surfaces, with hydrophilic triangular patches and superhydrophobic surrounding, impose direction-dependent adhesion. This dependency is quantified by bending beam experiments, in which a droplet is dragged on the surface in two opposite directions. In addition, force calculation derived from droplet roll-off experiments reveal accordant trends. The lower the aspect ratio, the higher the difference of adhesion forces between the two directions. The shape of the fluid contact line on the biphilic surfaces elucidates the origin of the direction-dependent adhesion. Namely, the droplet receding part pins to a higher number of patches when moving toward to apex in comparison to the opposite direction. When the triangular patches are more slender, this asymmetry cancels. These findings improve the understanding of droplet adhesion to biphilic surfaces particularly, and surfaces with wetting heterogeneities in general.

## Introduction

Adhesion and pinning forces of droplets to surfaces govern their motion and dynamics[1,2]. These transport phenomena are highly important for controlling fluids at interfaces, namely collection, removal, or accumulation of liquids at specific places[3]. These are highly relevant to heat transfer across interfaces[4], inkjet printing and surface patterning[5,6], and collecting condensed water in applications such as fog and dew harvesting[7]. The adhesion forces of droplets to surfaces are derived from the contact angle (CA) hysteresis[8,9], namely the difference between the advancing and receding contact angles of the moving droplet[10,11]. This lateral adhesion force is given in **Equation 1**:[12,13]

$$F_{Lateral} = k\gamma L_s(\cos\theta_{Rec} - \cos\theta_{Adv}) \qquad (1)$$

$F_{Lateral}$ is the lateral adhesion force, acting at the contact of the droplet with a solid surface, $k$- a pre-factor, $\gamma$- interfacial tension of the liquid, $L_s$- contact width, and $\theta_{Rec}$ and $\theta_{Adv}$ are the receding and advancing contact angles, respectively. The pre-factor $k$ usually equals 1 for simplicity.[14] This equation, however, does not account for surface asymmetries, namely when surface features are directional. Contact angle hysteresis can be caused by surface topography[15,16] and defects[17], often referred to as physical properties. It can also result from variations in surface wettability, derived from chemical heterogeneities[18]. The latter is the origin of contact angle hysteresis on flat surfaces. In recent years, the term "biphilicity" was coined to describe surfaces with hydrophilic patches on a hydrophobic background that are regular or meant to induce a specific function in liquid transport. Because of the excess of boundaries between the hydrophilic and hydrophobic areas, biphilic surfaces show high hysteresis. Therefore, studies aimed at finding optimal patch size and geometry on such surfaces in order to capture droplets, increase condensation rates, and control droplet transport on them[19]. A variety of surfaces in nature and engineering are biphilic[20–23]. In engineering, biphilic surfaces were investigated in terms of condensation[24] and evaporation[25], defrosting dynamics[4], pool boiling[26,27], and printing[6].

While the motion of fluids on solid surfaces has been studied for many years[13,28–30], it was recently experimentally shown that the adhesion of droplets to surfaces has the same behavior as friction between two solids[12]. The adhesion force is divided into static and dynamics regimes. The static adhesion force grows monotonically until it reaches a threshold, above which the droplets start sliding. Consequently, the adhesion force reduces to the kinetic force.

Previous studies investigated the stability of droplets on inclined surfaces in terms of roll-off angles[31] and observations of droplets contact lines with surfaces[5]. Directional transport of fluids on patterned surfaces was reported without direct quantification of forces and their dependence on specific directions[32,33]. Nevertheless, quantification of adhesion forces and droplet roll-off on biphilic surfaces remained mostly under studied. Moreover, direction dependent water transport due to asymmetric biphilicity was not addressed, although such findings are highly relevant to engineering and fundamental understanding of liquid transport in nature.

Here, we quantify the transport dynamics and adhesion forces of droplets on asymmetric biphilic surfaces. We focus on the underlying physics behind direction dependent adhesion of droplets, when these interact with surfaces with chemical heterogeneities. The biphilicity of the surfaces in this study is derived from triangular hydrophilic patches ($CA_{static}$ = 60 °± 10 °) on a superhydrophobic background ($CA_{static}$= 165 °± 2 °). We use two independent experimental techniques to calculate lateral adhesion forces of droplets to asymmetric biphilic surfaces: bending beam experiments and droplet roll-off on inclined surfaces.

In the bending beam experiments, a droplet is placed on the surface, on a horizontal motorized linear stage. The droplet is pinned onto a bending beam and small displacements of the beam are measured using an optical fiber. The lateral adhesion forces are calculated using bending beam theory. Second, a droplet is placed on a biphilic surface placed on a goniometer, and the adhesion force is derived from the gravitational force at roll-off. To understand the differences in adhesion force between the two directions of motion, we use fluorescence microscopy to elucidate the nature of contact between the droplets and the biphilic surfaces.

*Droplet motion over an asymmetric biphilic surface*

First, we calculate adhesion forces from measuring small deflections of a bending beam (**Figure 1**a). To convert the deflection of the beam to force, we use the Euler-Bernoulli beam theory for small deflections (**Equation 2**)[34]:

$$\delta_{max} = \frac{ML^3}{3EI}, I = \frac{WT^3}{12} \qquad (2)$$

with $\delta_{max}$- the deflection at the free end, $M$- bending moment, $L$- length of the beam, $E$- Young's modulus of the material, here $2 \cdot 10^5\ MPa$ for Steel 304, $I$- moment of inertia, $W$- width, and $T$- thickness of the beam.

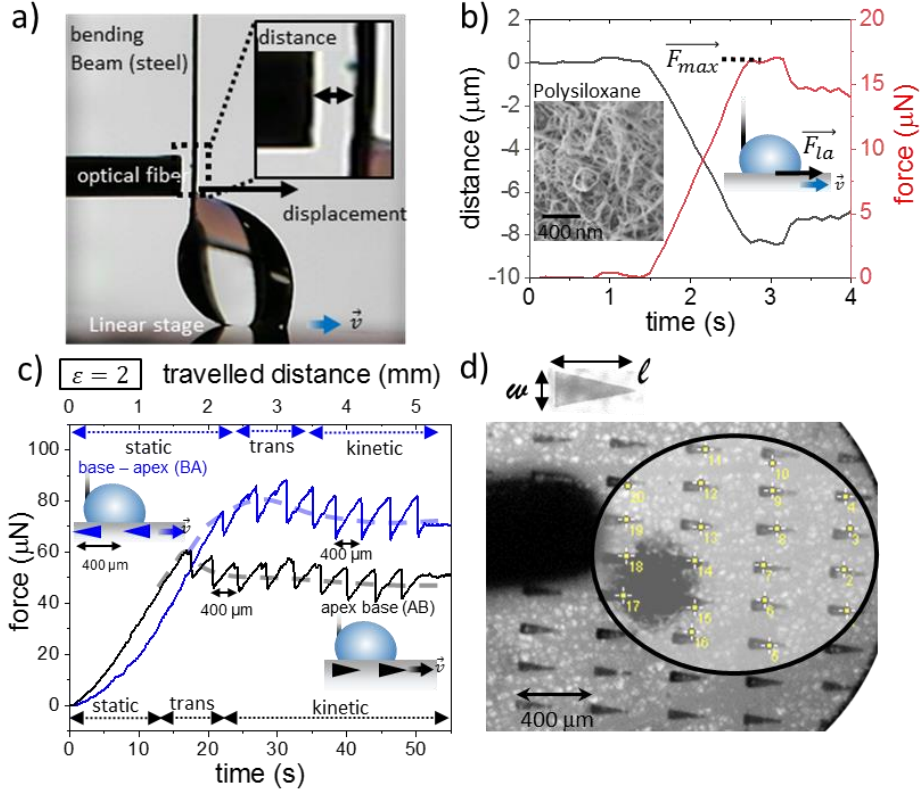

**Figure 1**. a) Bending beam setup for measuring lateral adhesion forces of droplets on different surfaces. A droplet is moved by a linear stage. The deflection of the beam is measured by an optical fiber. b) Measured displacement (black) and calculated lateral adhesion force $\vec{F}_{la}$ (red) of a water droplet on a superhydrophobic surface (inset). The droplet starts sliding when it overcomes the maximum static adhesion force $\vec{F}_{max}$. c) Lateral adhesion forces of a droplet (7.5 μl), sliding over a biphilic surface (triangular hydrophilic patches with aspect ratio $\varepsilon = 2$), with a velocity of 0.1 mm·s$^{-1}$, in two directions: AB- the droplet moves toward the base of the triangles (black), and BA-the opposite direction (blue). The dashed lines provide guide to the eye. d) Top view of the biphilic surface with a droplet on top (black circle). Patch aspect ratio corresponds to $\varepsilon = l/w$.

The displacement of a 10 μl water droplet, placed on a control superhydrophobic polysiloxane surface ($CA_{static}$= 165 °± 2 °), was measured (black) and the force calculated

(red), as shown in Figure 1b. The bending beam was inserted into the droplet and the stage moved with a velocity of 0.1 mm·s$^{-1}$. The beam shows zero displacement prior to the droplet motion. When the pulling force, exerted on the droplet by the beam, exceeds the static lateral adhesion force, the drop starts sliding on the surface. The lateral adhesion force drops and reaches the kinetic regime, in which it is roughly constant. The transition between the maximal force and the kinetic regime is called transitional regime[12]. The average of the maximal static adhesion forces of the droplet on the superhydrophobic surface was $\bar{F}_{static}^{max} = 17 \pm 3$ µN. This value agrees with previous measurements on superhydrophobic surfaces,[12,35] and was independent of the direction of motion.

We then introduce hydrophilic isosceles triangular patches to the superhydrophobic surface and measure the adhesion force of a droplet to the biphilic surface (Figure 1c). The triangles have a length of $l = 200$ µm and a width of $w = 100$ µm ($\varepsilon = l/w = 2$). A 7.5 µl water droplet was placed on the substrate, the bending beam was immersed into the drop and the linear stage moved at a velocity of 0.1 mm·s$^{-1}$.

Droplets moving over periodic symmetric structures show equidistant peaks in force measurements, similar to the periodicity of the structure.[35] The peaks result from de-pinning of the receding side of the drop. We observe a chainsaw shaped force curve that fits the spacing between the hydrophilic patches (Figure 1c). The sample was placed parallel to the bisecting line, connecting the apex and the base center. The lateral adhesion of the droplet to the surface was measured in two directions of motion: i) the droplet passes first the base and leaves the triangle at the apex, namely BA direction (blue), and ii) the opposite direction with the apex first and base second, namely AB direction (black). The droplet covers multiple patches, depending on its size and placement. For example, a 5 µl droplet covers roughly 20 patches (Figure 1d). When the stage starts moving, the adhesion static force increases until it reaches the maximal static adhesion force. Further movement of the stage causes the droplet to start sliding. After the maximal force is reached, a drop in the adhesion force is observed, noted as the transitional regime. This is a result of the depinning of the contact line at the rear side of the droplet (Supporting information, **Video S1**). The average maximal static adhesion force in BA direction corresponds to $\bar{F}_{static}^{max\,BA} = 76 \pm 13$ µN, while in the opposite direction, AB, this force corresponds to $\bar{F}_{static}^{max\,AB} = 57 \pm 5$ µN. Subsequently, the drop enters the kinetic regime, in which the droplet motion is characterized by a synchronized jumping of the front and back of the droplet, reflected in force peaks with a distance of 400 µm in between. This is the distance between the bases of two adjacent triangles along the angle bisector line. The difference between the maximal and minimal forces during the stick-and-slip kinetic regime ranges between 10 and 15 µN in BA direction. In AB direction, a similar type of chainsaw force curve is observed, with force differences of 8 to 10 µN. The average kinetic adhesion force in BA direction is $\bar{F}_{kinetic}^{BA} = 70 \pm 12$ µN, and $\bar{F}_{kinetic}^{AB} = 37 \pm 10$ µN in AB direction. Both in the static and kinetic regimes, the forces in AB direction are roughly two-third the forces in the BA direction. In some cases, there are small force peaks of roughly 2 µN, resulting from local depinning of the contact line from single patches (see Supporting Information, **Figure S1**). Continuous motion of the droplet on the biphilic surface is characterized by a cyclic stick-slip behavior. Namely, each de-pinning event results in a jump of the droplet contact line to the next row of patches. Upon de-pinning, a thin visible film of water remains on the patches (see Supporting Information, **Figure S2**).

*The role of gradients in direction-dependent adhesion*

We now examine the influence of the triangular patches' aspect ratio on the direction-dependent adhesion force (**Figure 2**). The length-to-width ratio, ε, was varied between ε = 2 (Figure 2a) and $\varepsilon = 3.7$ (Figure 2b). The forces in AB direction appear black and in BA direction appear blue. The dashed lines provide guides to the eye. We observe differences in the direction dependent adhesion between the two surfaces, both in the static and kinetic regimes. To quantify these differences, we discuss two quantities. First, the difference between the maximal static adhesion forces on each biphilic surface, given by **Equation 3**:

$$\Delta F_{static}^{max} = F_{static}^{max\,BA} - F_{static}^{max\,AB} \qquad (3)$$

With $F_{static}^{max\,BA}$ and $F_{static}^{max\,AB}$ the maximal static adhesion forces in BA and AB directions, respectively. Second, we quantify the direction dependent difference between the average adhesion forces in the kinetic regime, given by **Equation 4**:

$$\Delta \bar{F}_{kinetic} = \bar{F}_{kinetic}^{BA} - \bar{F}_{kinetic}^{AB} \qquad (4)$$

With $\bar{F}_{kinetic}^{BA}$ and $\bar{F}_{kinetic}^{AB}$ the average adhesion forces in the kinetic regime in BA and AB directions, respectively.

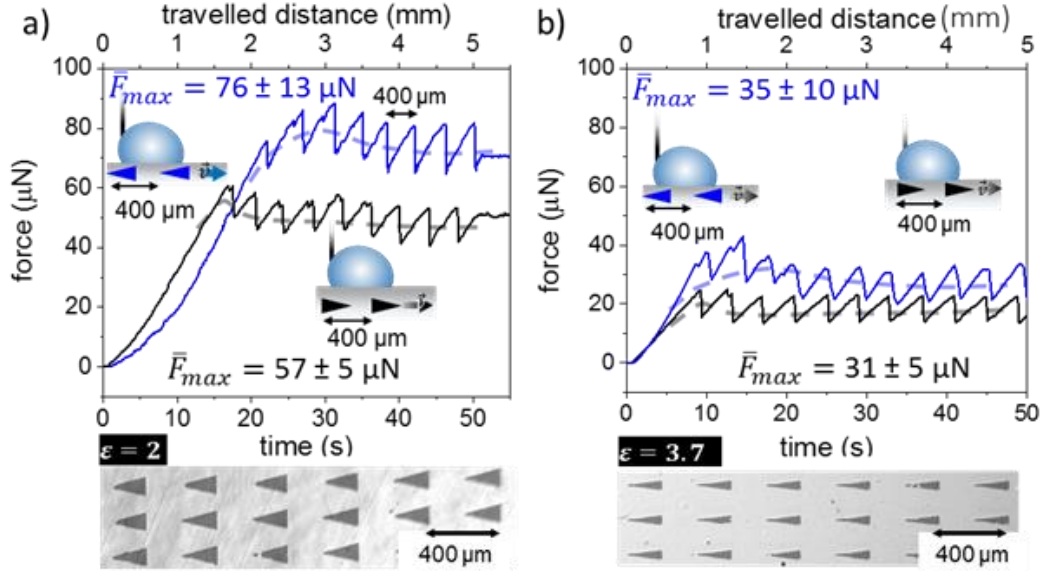

**Figure 2**. Representative lateral adhesion force measurements in AB (black) and BA (blue) directions on biphilic surfaces with triangular patches of length-to-width ratio of a) $\varepsilon = 2$, and b) $\varepsilon = 3.7$. The triangle edge corresponds to 200 µm in all cases. The droplet volume is 7.5 µl.

$\Delta F_{static}^{max} = 19 \pm 8$ µN for $\varepsilon = 2$ and $\Delta F_{static}^{max} = 3 \pm 3$ µN for $\varepsilon = 3.7$. The average forces, in the kinetic regime, for $\varepsilon = 2$ correspond to $\bar{F}_{kinetic}^{BA} = 70 \pm 12$ µN in BA direction and $\bar{F}_{kinetic}^{AB} = 37 \pm 10$ µN in AB direction. This results in a force difference of approximately $\Delta \bar{F}_{kinetic} = 33$ µN. For $\varepsilon = 3.7$, the force difference is $\bar{F}_{kinetic}^{BA} = 27 \pm 12$ µN in BA direction, and $\bar{F}_{kinetic}^{AB} = 15 \pm 5$ µN in AB direction. This results in a force difference of $\Delta \bar{F}_{kinetic} = 12$ µN. Overall, the forces are consistently higher for BA direction in all regimes, both for $\varepsilon = 2$ and $\varepsilon = 3.7$. However, the differences in forces become smaller for a higher aspect ratio, in which the geometrical gradient is reduced. Specifically, the difference between the average forces in the kinetic regime in AB and BA directions is cut by half for patches with aspect ratio $\varepsilon = 3.7$, in comparison to $\varepsilon = 2$.

*Droplet roll-off on asymmetric biphilic surfaces*

Next, we quantify the adhesion forces of the droplet to biphilic surfaces, by roll-off experiments (**Figure 3**). A droplet of a defined volume was placed on the surface and the stage was tilted until roll-off. The sample was turned 90 ° and the droplet was placed on the same location to ensure similar surface conditions. The force the droplet overcomes in order to roll-off corresponds to the force $F_{static}^{max}$, measured using the bending beam[36].

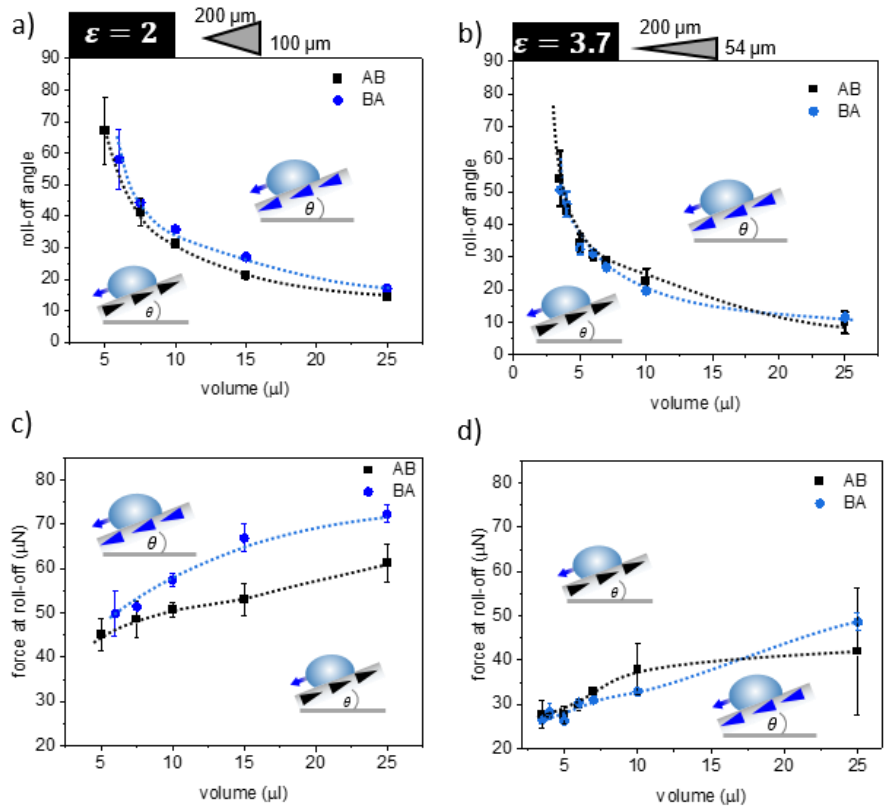

**Figure 3.** Roll-off angles and corresponding lateral adhesion forces at roll-off, depending on droplet volume, on biphilic surfaces with triangular isosceles hydrophilic patches. The patch length-to-width ratio corresponds to $\varepsilon$= 2 (a, c) and $\varepsilon$= 3.7 (b, d).

Droplet roll-off occurs when the gravitational force surmounts the adhesion or pinning forces. Namely, when the difference between the dynamic advancing and receding contact angles overcomes the contact angle hysteresis[29]. The droplet shape deforms due to the gravitational force (supporting information, **Figure S3**). Therefore, the advancing contact angle increases, approaching 180 °,[37] and the receding contact angle decreases.

Droplet roll-off, depending on droplet volume, was measured for the surfaces with $\varepsilon = 2$ and 3.7 in BA and AB directions. An increase in drop volume leads to a monotonic decrease in roll-off angle, first rapidly and consequently asymptotically. Within this range of volumes, the transition between the capillary regime and gravitational regime takes place. The Bond number, $Bo = \left(\frac{L}{\lambda_c}\right)^2$, represents the ratio between gravitational and capillary forces, with $L$- characteristic length and $\lambda_c$ – capillary number. Considering $L$ as the diameter of the droplet, the bond number increases from 0.25 for a droplet volume of 3.5 µl to 1.1 for 25 µl. Roll-off initiates when gravity overcomes lateral adhesion forces, and therefore, the latter can be calculated from measuring the roll-off angle, as shown in **Equation 5**.[17,29,30]

$$F = mg \sin(\theta_r) \quad (5)$$

Where $m$ is the mass of the droplet, $g$- gravitational acceleration and $\theta_r$- roll-off angle. In detail, for $\varepsilon = 2$, motion in AB direction shows a lower roll-off angle (Figure 3a). The calculated force in AB direction, necessary to initiate motion, is lower compared to BA direction for all droplet volumes. The forces in AB vary from 45 µN to 60 µN for droplet volumes of 5 µl to 25 µl, respectively. These forces are in the same order of magnitude as the lateral adhesion forces measured using the bending beam. Specifically, for a volume of 7.5 µl, a force of $F_{static}^{\max AB} = 48 \pm 4$µN is calculated from roll-off, compared to $F_{static}^{\max AB} = 57 \pm 5$ µN, measured by the bending beam. In BA direction, for a 7.5 µl droplet, a force of $F_{static}^{\max BA} = 51 \pm 2$µN was calculated from roll-off, compared to $F_{static}^{\max BA} = 76 \pm 14$ µN, using the bending beam. Roll-off occurs at lower forces in comparison to the adhesion forces obtained by the bending beam. This difference may result from mechanical vibrations of the goniometer, inducing droplet roll-off earlier.

Roll-off on surfaces with $\varepsilon = 3.7$ shows smaller variations in the roll-off angles between the two directions. In BA direction, the force difference ranges from $\Delta F_{static}^{BA\,max} = 26 \pm 1$ µN, for 3.5 µl, to $\Delta F_{static}^{BA\,max} = 48 \pm 1$ µN, for 25 µl. In AB direction, the force difference ranges from $\Delta F_{static}^{AB\,max} = 28 \pm 2$ µN, for 3.5 µl, to $\Delta F_{static}^{AB\,max} = 41 \pm 14$ µN, for 25 µl. Forces for roll-off, namely the maximal static forces, of 7 µl droplets were measured to be

$F_{static}^{BA\ max} = 31 \pm 1$ µN and $F_{static}^{AB\ max} = 33 \pm 1$ µN, for BA and AB directions, respectively. These values coincide with the forces measured in the bending beam experiments with $F_{static}^{BA\ max} = 35 \pm 10$ µN and $F_{static}^{AB\ max} = 31 \pm 5$ µN.

*Asymmetry of contact line pinning on directional biphilic surfaces*

We expect the receding part of droplets to reveal the underlying mechanism for the direction dependent adhesion described above. To this end, we visualize the motion of droplets on asymmetric biphilic surfaces using fluorescence microscopy (**Figure 4**).

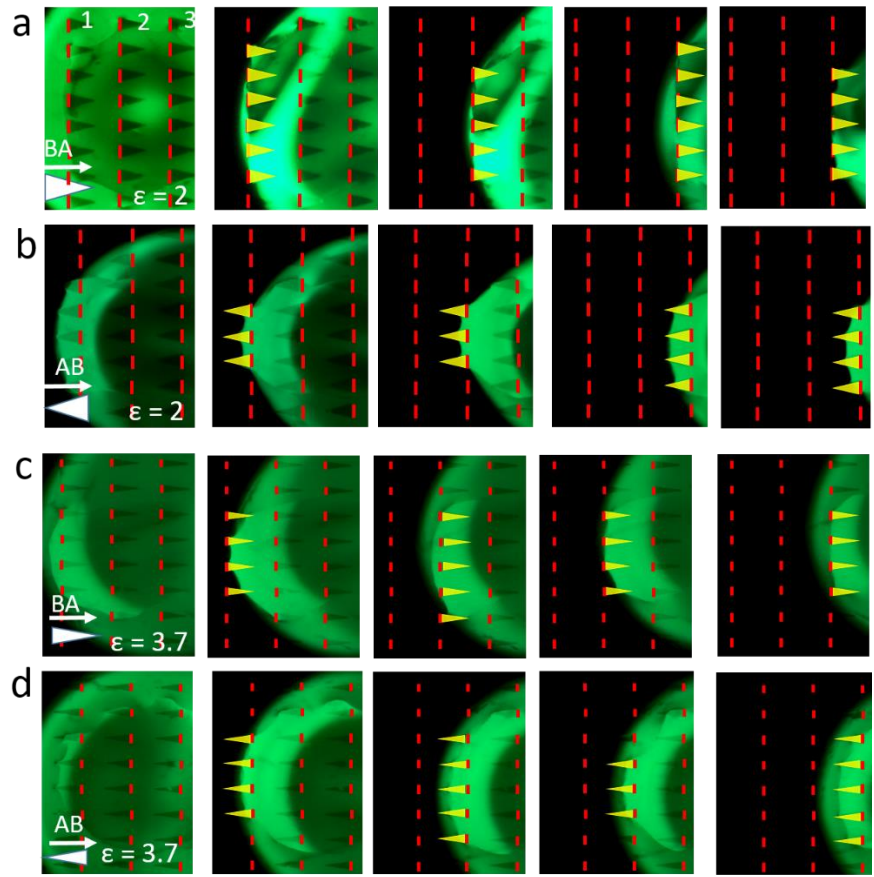

**Figure 4**. Receding contact lines of moving droplets in two opposite directions for aspect ratio $\varepsilon = 2$ (a,b) and $\varepsilon = 3.7$ (c,d). Epi-fluorescent microscopy of a 7.5 µl droplet dragged along a biphilic surface in direction BA (a,c) and AB (b,d), corresponding to base-apex and apex-base directions, respectively, indicated by white arrows. The red dashed lines mark the bases of the patches. The pinning patches, shortly before release of the receding contact line, are marked with yellow triangles for better visualization.

As shown in the preceding sections, the maximal adhesion force depends on two factors: a) the asymmetry of the patches, i.e. the length-to-width ratio, $\varepsilon$, and b) the direction of the droplet motion in respect to the patch. Epi-fluorescence imaging shows the contact line of a sliding droplet over asymmetric biphilic surfaces. Snapshots of the receding contact line depict droplets before jumping from a line of patches to the consecutive line (marked with red dashed lines, numbering 1, 2, 3). Patches to which the droplet is pinned are marked with yellow triangles. We consider first a droplet moving over patches with an aspect ratio of $\varepsilon = 2$. In BA direction, the droplet pins to approximately 5 to 6 patches. In the opposite direction, however, pinning occurs only on 3 to 4 patches. This difference, of approximately 2-3 patches in each depinning event, is likely to cause larger pinning forces in BA direction. This results in an averaged difference of $\Delta \bar{F}_{kinetic} = 33$ µN between the two directions of motion in the kinetic regime. The contact line is released mostly from the base of the triangles, independent of the direction. When asymmetry is reduced, for length-to-width ratio of $\varepsilon = 3.7$, pinning becomes more uniform and independent of direction. Namely, the difference between the number of patches involved in each depinning event, depending on the direction of motion, varies between 0 and 1. This observation is consistent with the force measurements, showing a smaller difference of $\Delta \bar{F}_{kinetic} = 12$ µN between the adhesion forces in the kinetic regime in both directions. The length of the contact line, in contact with the hydrophilic patches, determines the maximal adhesion forces. As the droplet easily de-pins from the hydrophobic background the adhesion force to overcome for forward motion depends on the number of the pinning patches. Specifically, the larger the number of patches, the larger the pinning. Equation 1 gives the retentive force a droplet needs to overcome in order to move. The retentive force depends on a) the width and b) shape of the droplet. In our study, the width and shape of the droplet depend on the direction of motion, as indicated by the number of patches pinning droplets on the receding part in each direction. Since for $\varepsilon = 2$ in BA direction more patches pin the receding part in comparison to the AB direction, the width of the droplet is expected to be larger in BA. In addition, the contact line of droplets interacting with the patches of $\varepsilon = 2$ is longer than for $\varepsilon = 3.7$. Therefore, the contact angle hysteresis is larger due to the larger area fraction of the hydrophilic patches, in comparison to $\varepsilon = 3.7$. The area fraction for $\varepsilon =$

2 is 0.25, almost double the area fraction of $\varepsilon = 3.7$, corresponding to 0.135. This explains the lower adhesion on the patches with $\varepsilon = 3.7$, in comparison to $\varepsilon = 2$.

*Summary and conclusion*

In summary, we show a stick-and-slip motion of droplets on asymmetric biphilic surfaces. The droplet three-phase contact line jumps along rows of triangular hydrophilic patches. The maximal static adhesion force and the average adhesion force in the kinetic regime depend on a) the direction the droplet motion and b) the length-to-width ratio of the patches. The force is consistently lower for all length-to-width ratios in AB direction, in which the droplet passes first the apex and then the base. The maximal static adhesion force on biphilic surfaces with patch of $\varepsilon = 2$ differs in roughly 20 µN between the opposite directions of motion. In AB direction, the average kinetic force corresponds to 57 µN, whereas in BA direction to 76 µN. For slender triangular patches, with $\varepsilon = 3.7$, the difference in adhesion forces between the two directions almost cancels. Maximal static adhesion forces, calculated independently by roll-off angle measurements and a bending beam, show consistent results.

A deeper insight of the contact line at the receding part of the sliding droplet reveals the underlying mechanism of the direction and geometry dependent adhesion. The direction dependence of the adhesion forces for $\varepsilon = 2$ is a result of the patch coverage and pinning, which is lower in AB direction. Therefore, the droplet is subjected to lower adhesion forces. This asymmetry almost cancels for slender triangles with $\varepsilon = 3.7$. Larger patch area fraction, corresponding to smaller length-to-width ratios, increases the contact angle hysteresis and leads to an increase in adhesion force. Alternatively, droplets move more easily toward the base of triangular patches. The direction dependent adhesion is sensitive to the gradients in the patch geometry, namely, the length-to-width ratio. These results set guidelines for designing biphilic surfaces on which droplets are expected to roll-off or to be collected in specific areas.

**Materials and Methods**

*Sample preparation*

Pre-cut silicon wafers or glass slides were cleaned in base piranha solution (5:1:1 $H_2O$:ammonium hydroxide:hydrogen peroxide) heated to 75 °C for 15 minutes and rinsed with water three times and dried in the oven at 110 °C for 5 min. Oxygen plasma (Diener electronic GmbH + Co. KG) was applied for 2 min. The cleaned surfaces were further treated for gaining superhydrophobicity[38]: the samples were placed in a disposable vial with Toluene. Per each 10 ml of Toluene 95 μl methyltrichlorosilane and 250 μl were added. Over a reaction period of 3h a polysiloxane network grew consisting of cylindrical fibers with ~40 nm diameter (Figure 1b, inset). The final steps were washing the samples in Toluene, Ethanol, Ethanol-water 50:50 mixtures, and water 3 times respectively, followed by tempering at 110 °C in the oven. This resulted in samples with a superhydrophobic character i.e. static contact angles of 165 °± 2 ° and roll-off angles of approximately 2°. The samples were ready for further modification or use. 2D shadow masks were prepared by laser cutter (The ELAS Master Femto, ELAS Ltd. Lithuania) in commercially available plastic foil for overhead projectors. A hydrophilic pattern was created by a magnetron sputterer (PVD20, Vinci Tech., France) of titanium through a shadow mask onto a substrate such as a superhydrophobic surface with thicknesses of about 30 nm. The measured contact angle after Ti deposition is 60° ± 10°.

*Contact angle measurements and goniometer*

The setup for roll-off measurements is based on a contact angle measurement system OCA25 (DataPhysics Instruments GmbH, Germany) with a tilting base (0 ° to 95 °). A side view camera for contact angle measurements was accompanied by a top view camera (Fastcam Mini AX50, Photron USA Inc.) and a lens system (Infinity, USA).

*Lateral adhesion force measurements*

Measurements were performed in a setup of a fiber optic displacement sensor, a bending beam, and a linear stage (**Figure 1a**). The bending beam (steel 304, thickness 0.09 mm, length 30 mm, width 1.45 mm) was fixed at one end and the free end was placed into a droplet that was placed onto the surface of interest. The motorized linear stage ( LTS 150,

Thorlabs) moved the sample at a velocity of 0.1 mm/s if not mentioned otherwise. When the bending beam reached the edge of the droplet the droplet was dragged over the surface. Due to the lateral adhesion forces the beam bends. The deflection of the beam was measured by a fiber optic sensor (muDMS RC20, Philtec Inc. USA). In all the experiments milli-Q water (resistivity 18.2MΩ cm) was used.

*Fluorescence Microscopy*

Epi-fluorescence images were recorded using an inverted microscope Axiovert A1 with a color camera Axiovam 208 (Carl Zeiss Microscopy GmbH, Germany) and transparent modified glass samples. The camera recorded with 5 fps.

**Supporting Information**

**Video S1**

A 7.5 µl droplet is dragged over a biphilic surface with triangular patches with a length-to-width ratio of 2 (width = 100 µm, length = 200 µm). The distance between patches corresponds to 200 µm.

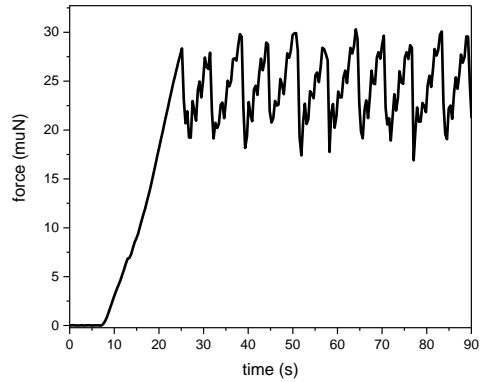

**Figure S1.** Adhesion force on a biphilic surface with ε = 3.7 at a velocity of 1 mm/s.

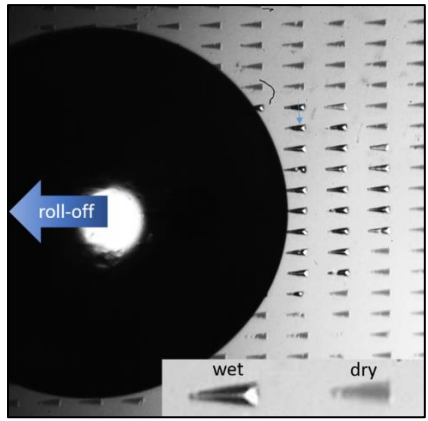

**Figure S2.** Residual patch wetting after droplet roll-off.

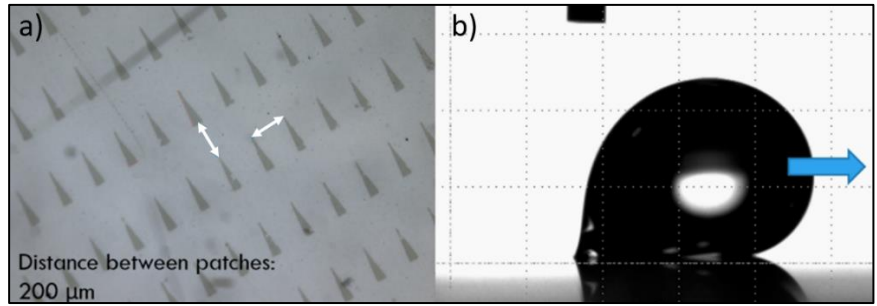

**Figure S3**. a) Micrograph of a biphilic surface with triangular hydrophilic patches made of titanium on a hydrophobic background (l/w = 3.7). The white arrows represent the distance of 200 μm between the patches. b) Droplet roll-off from a biphilic patch displayed from a side view camera.

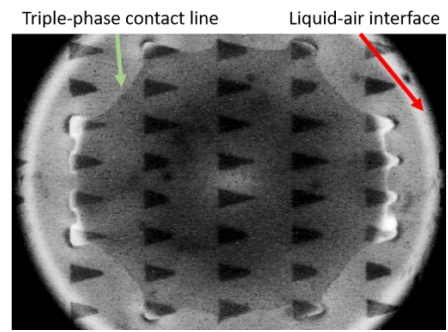

**Figure S4**. Triple phase contact line of a 7.5μl droplet on a biphilic surface with $\varepsilon = 2$.